\documentclass[twocolumn,aps]{revtex4}
\usepackage{amsmath}
\usepackage{amssymb}
\usepackage{graphicx}
\usepackage{dcolumn}
\usepackage{bm}
\usepackage{xcolor}
\usepackage{epstopdf}

\begin{document}

\preprint{Phys.Rev.B}

\title{Viscous magnetotransport and Gurzhi effect in bilayer electron system}
\author{G. M. Gusev,$^1$
A. S. Jaroshevich,$^{2}$  A. D. Levin,$^1$  Z. D. Kvon,$^{2,3}$
 and A. K. Bakarov $^{2,3}$}

\affiliation{$^1$Instituto de F\'{\i}sica da Universidade de S\~ao
Paulo, 135960-170, S\~ao Paulo, SP, Brazil}
\affiliation{$^2$Institute of Semiconductor Physics, Novosibirsk
630090, Russia}
\affiliation{$^3$Novosibirsk State University, Novosibirsk 630090,
Russia}

\date{\today}
\begin{abstract}
We observe a large negative magnetoresistance and a decrease of resistivity with increasing temperature, known as the Gurzhi effect,
in a bilayer electron (BL) system formed by a wide GaAs quantum well. A hydrodynamic model for
the single fluid transport parameters in narrow channels is employed and successfully describes our experimental findings.
 We find that the electron-electron scattering in the bilayer is more intensive in comparison with a single-band well (SW).
 The hydrodynamic assumption implies a strong dependence on boundary conditions,
which can be characterized by slip length, describing the behavior of a liquid near the edge.
Our results reveal that slip length in a BL is shorter than in a SW,
and that the BL system goes deeper into the hydrodynamic regime. This is in agreement with the model proposed where the slip length is of
the order of the electron-electron mean free path.

\end{abstract}

\maketitle

\section{Introduction}
Transport of finite size conductors is strongly affected by electron-electron interactions.
In dissipative hydrodynamic phenomena, the key physical parameters, controlling the Poiseuille-like flow profile, are shear stress relaxation time $\tau_{ee}$ due to electron electron collisions and the slip length $l_{s}$, which characterizes the behavior of a liquid near the edge (Fig.1). The hydrodynamic regime requires $l/l_{e} \gg  1$ and  $l_{e}/W\ll 1$, where $l$ is the mean free path of electrons with respect to momentum changing scattering  by impurities  and  phonons, $W$ is  the  channel  width, and $l_{e}$ is the mean  free path for electron-electron collision \cite{gurzhi}-\cite{raichev}. In addition,
while the flow is Poiseuille-like for $l_{s}/W \ll 1$, it becomes Ohmic for $l_{s}/W \gg 1$ \cite{kiselev, pellegrino2}.

The most prominent manifestation of electron-electron interaction associated with hydrodynamic electron flow
has been predicted in the pioneering theoretical study by Gurzhi \cite{gurzhi}. It has been shown that resistance decreases with the
square of temperature,
$\rho \sim l_{e} \sim T^{-2}$ and with the square of sample width
$\rho \sim W^{-2}$. The Gurzhi effect is in apparent contradiction
with semiclassical transport theory because it results
in a decrease in the electrical resistivity, where collisions
become more frequent. In a two-dimensional (2D) system, a temperature-induced decrease of differential resistivity due to heating by the current
has been observed in GaAs wires \cite{dejong}, and
 a decrease of low current linear resistivity with T in H-shaped bar geometry samples \cite{gusev1}, both attributed to the Gurzhi effect.
 Many other features related to electron viscosity in  2D systems  have
 been  found  in  the  presence  of  a  magnetic field \cite{alekseev1,scaffidi,gusev1,gusev2,haug,hatke,mani,haug2,shi}.
\begin{figure}[ht]
\includegraphics[width=8cm]{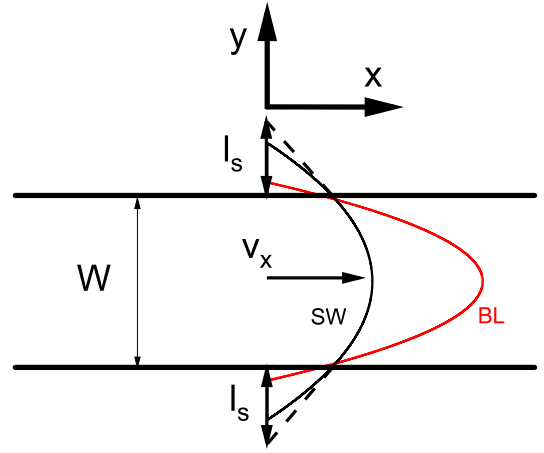}
\caption{(Color online) Schematic illustration of the flow profiles in a channel of width W for a single well and bilayer. The slip length $l_{s}$  corresponds to the length where the
extrapolated velocity (dashes) vanishes. With growing slip
length, the flow profile becomes flat, i.e., more similar to Ohmic flow.  }
\end{figure}

The parameter that is more difficult to control is the slip length. It is expected that, in GaAs material, the slip length is dependent on the etching technique,
however, more insight into this issue has shown that, in both diffusive and
nearly specular boundary scattering conditions, $l_{s}$ depends on electron-electron scattering length \cite{kiselev, pellegrino2}.
In particular, it has been shown that  $l_{s}=\alpha l_{e}$,
 where $\alpha\approx 1$ for diffusive boundary scattering and $\alpha<< 1$ for specular scattering \cite{kiselev}. For this reason, studies of systems
 with more intensive e-e scattering that differ from standard Fermi gas and liquid are of crucial interest.

 In a bilayer system, electrons occupy two closely spaced subbands and an additional channel for scattering - intersubband scattering - is opened up.
Moreover, the doubling of the phase space for the intrasubband rate and more effective screening for interaction may reduce the electron-electron scattering length
and improve boundary conditions for hydrodynamic flow.

In the present paper, we study magnetotransport in narrow channels fabricated from a high-quality bilayer electron system in a wide quantum well.
Owing to the charge redistribution, a wide well forms a bilayer electron configuration, where the two wells are separated by an
electrostatic potential barrier ( Fig.2, top) \cite{wiedmann} near the interfaces. We observe the resistance drop in a wide interval of temperature, consistent with the
prediction of viscous transport in narrow channels. In the presence of a transverse magnetic field, the samples display a Lorentzian-shaped
magnetoresistance in good agreement with magnetohydrodynamic theory \cite{alekseev1}.

\section{Experimental results: macroscopic sample}
 Our samples are high-quality, GaAs quantum wells with a width of 46 nm
 with electron density $n_{s}=n_{total}=6.7\times10^{11} cm ^{-2}$ and a mobility of $\mu=2\times10^{6} cm^{2}/Vs$ at T=1.4K.
 The charge distribution in a wide single quantum well is more subtle than the one in a double quantum well.
 Here the Coulomb repulsion of the electrons in the well leads to a soft barrier inside the well, which in turn results in a bilayer electron system.
 The calculated confinement potential profile of our wide quantum wells and electron wave functions for the first two subbands is shown
 in Figure 2a. The small energy separation and the symmetry of the wave functions for the two lowest subbands show that corresponding
 (symmetric and antisymmetric) states are formed as a result of tunnel hybridization $\Delta$ of the states in the two quantum wells near the interfaces.
\begin{figure}[ht]
\includegraphics[width=8cm]{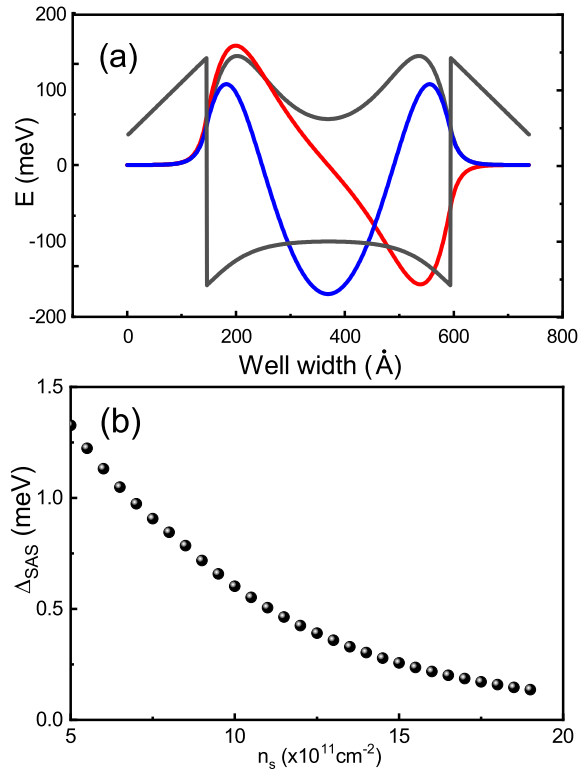}
\caption{(Color online)
(a) WQW with well width of 46 nm and corresponding symmetric (blue) and asymmetric (red) wave functions.
(b) Intersubband energy splitting as a function of well density.  }
\end{figure}
Figure 2b shows the dependence of the  $\Delta$ on total electron density. One can see that the energy separation drops with
density. This value of intersubband separation is close to 1 meV at density $6.7\times10^{11} cm^{-2}$ .

 \begin{figure}[ht]
 \includegraphics[width=8cm]{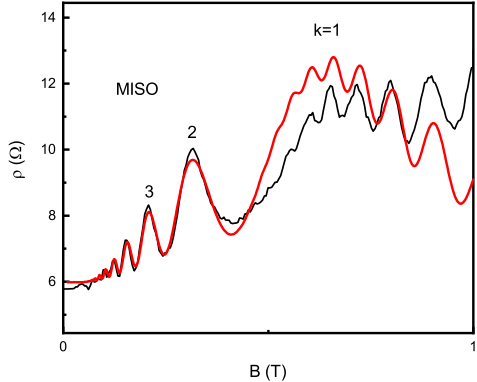}
\caption{(Color online)
Magnetointersubband oscillations  (MISO) and Shubnikov de Haas  oscillations of resistance in wide macroscopic QW samples, $T=1.5 K$.
 Comparison of the measured magnetoresistance with the calculated one (red line) allows us to determine subband separation $\Delta=1.1 meV$ by the oscillation frequency. }
\end{figure}

\begin{figure}[ht]
\includegraphics[width=10cm]{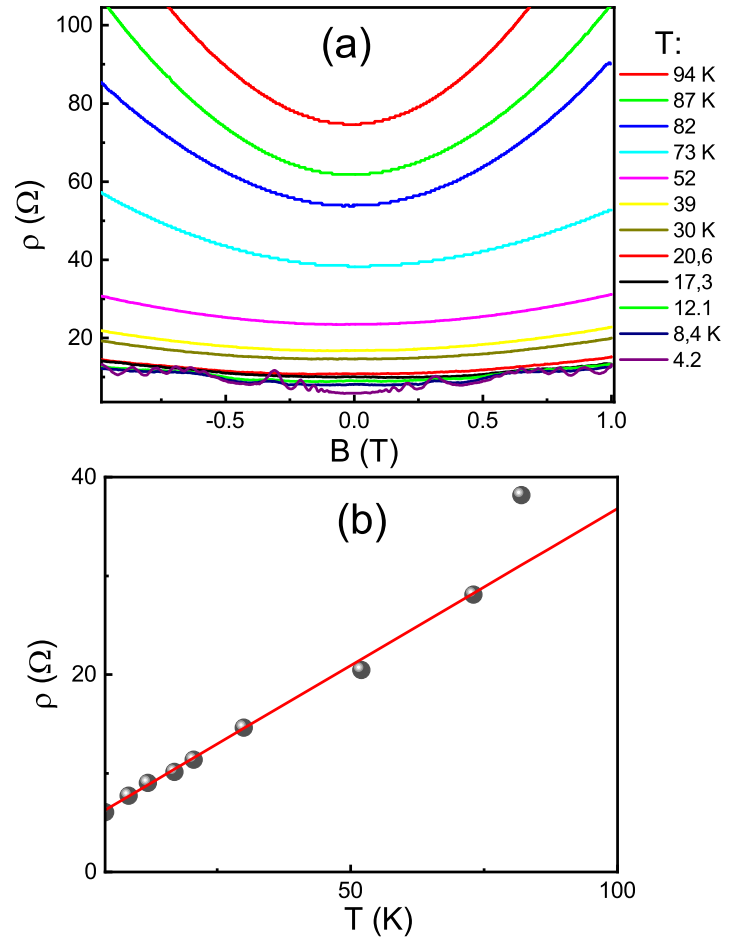}
\caption{(Color online)
(a) Temperature dependent magnetoresistance of a macroscopic GaAs quantum well.
(b) The linear temperature dependence of resistance at B=0 in macroscopic 2D samples.
 The red line is dependence : $\rho (T)/\rho (4.2) = 1+ \beta T$, with $\beta =0.07 K^{-1}$. }
\end{figure}
In quantum wells with two occupied 2D subbands, the magneto-resistance exhibits  oscillating behavior in  a magnetic field below where the conventional
Shunikov de Haas oscillations appear - the so-called
magneto-inter-subband (MIS) oscillations \cite{mamani, wiedmann}. These oscillations arise from the periodic modulation of the probability of transitions
between the Landau levels (LL) belonging to different subbands. The MIS oscillation is periodic in $\Delta/\hbar \omega_{c}$, where  $\omega_{c}=eB/mc$ is the cyclotron frequency.
Since the origin of the MIS oscillations is related to the alignment between the different Landau levels (LL) of the two subbands and
not to the position of the LL with respect to Fermi energy, these oscillations survive at high temperatures when SdH oscillations are suppressed.

Figure 3 shows magneto-resistance as a function of the magnetic field at $T=1.5 K$. The magneto-resistance exhibits MIS oscillations together with SdH oscillations.
The red line shows the theoretical MISO, which allows us to determine subband separation $\Delta=1.1 meV$ by the oscillation frequency.

 Figure 4(a) shows the longitudinal magnetoresistivity $\rho (B)$ measured in the local
configuration for a macroscopic  Hall bar sample ($500 \mu m\times 200 \mu m$) as a function of magnetic field and temperature. One can see an increase in zero field resistivity
$\rho_{0}$ with increasing T (Fig. 4(b)), and a positive parabolic magneoresistance.

Assuming the viscosity effect is small in macroscopic samples, we are able to fit an $\rho(T)$ dependence by the linear line:
$\rho(T)/\rho(4.2) = 1+ \beta T$ above 4.2 K, with $\beta =0.07 K^{-1}$ due to the contribution of electron-phonon scattering into the transport.

\section{Experimental results: mesoscopic sample}

We present experimental results on mesoscopic Hall-bar devices.
They consists of three, $6 \mu m$ wide consecutive segments of
different length ($6, 20 , 6 \mu m$), and 8 voltage probes.
The measurements were carried out in a
VTI cryostat, using a conventional
lock-in technique to measure the longitudinal $\rho_{xx}$ resistivity with an
ac current of $0.1 - 1 \mu A$ through the sample, which is
sufficiently low to avoid overheating effects.
2 Hall bars from the same wafers were studied and showed consistent behaviour.

\begin{table}[ht]
\caption{\label{tab2} Parameters of the electron system in mesoscopic samples at $T=1.4 K$. The mean free paths  $l=v_{F}\tau_{macrosc}$.
 Other parameters are defined in the text.}
\begin{ruledtabular}
\begin{tabular}{lcccccccc}
&W&$n_{s}$ & $v_{F}$ & $l$  & $l_{2}$ & $\eta$ & Properties\\
&$\mu m$ & $(10^{11} cm^{2}$) & $(10^{7} cm/s)$ &  $(\mu m$) & $(\mu m$) & $(m^{2}/s)$ & & \\
\hline
&$6$& $6.7$  & $2.5$ &  $19$ & $0.45$ & $0.07$ & BL\\
&$5$& $7.4$  & $3.7$ &  $28$ & $2.5$ & $0.23$ & SW\\
\end{tabular}
\end{ruledtabular}
\end{table}

\begin{figure}[ht]
\includegraphics[width=8cm]{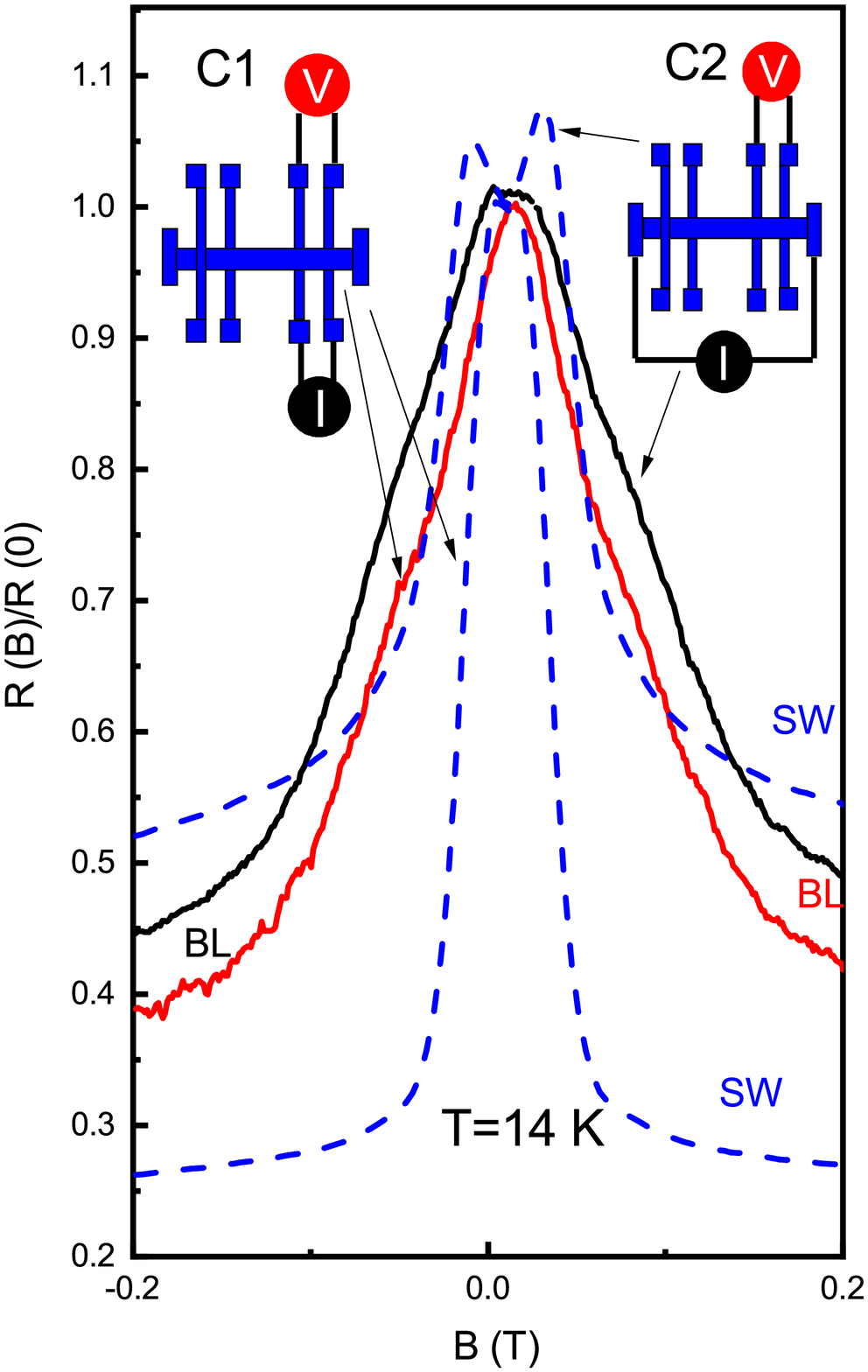}
\caption{(Color online) Longitudinal resistance in the 6-$\mu m$-wide mesoscopic Hall bar for different configurations of measurements
and for different devices fabricated from a single subband GaAs well (SW) (dashes) and a wide GaAs well or bilayer (BL) (solid lines), T=14 K. }
\end{figure}
The temperature dependence of resistance at B=0 in macroscopic 2D samples  was linear,
 with the coefficient given in previous section.
 For these parameters, the mean free path $l$ is larger than $W$  even at $T \sim 30 K$.
 The parameters characterizing the electron system are given in Table 1. For comparison we also
show parameters for one of the typical single well samples studied previously \cite{raichev}.

\begin{figure}[ht]
\includegraphics[width=9cm]{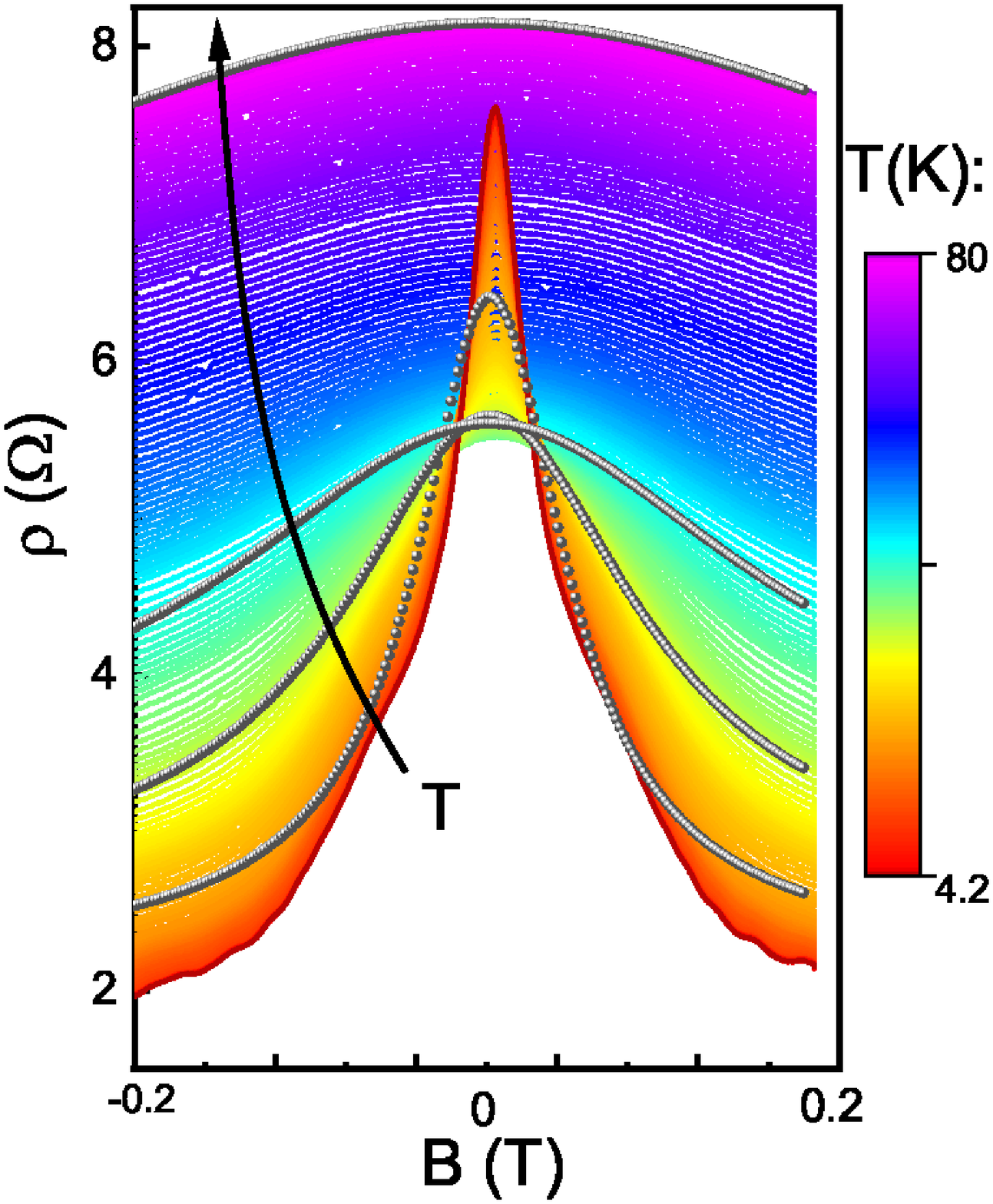}
\caption{(Color online)
Temperature-dependent  magnetoresistance  of  a mesoscopic GaAs bilayer for configuration C1. The circles are examples illustrating magnetoresistance calculated from
 Eq. (2) for different temperatures T(K): 14, 27, 43.9, 80.}
 
\end{figure}

\begin{figure}[ht]
\includegraphics[width=9cm]{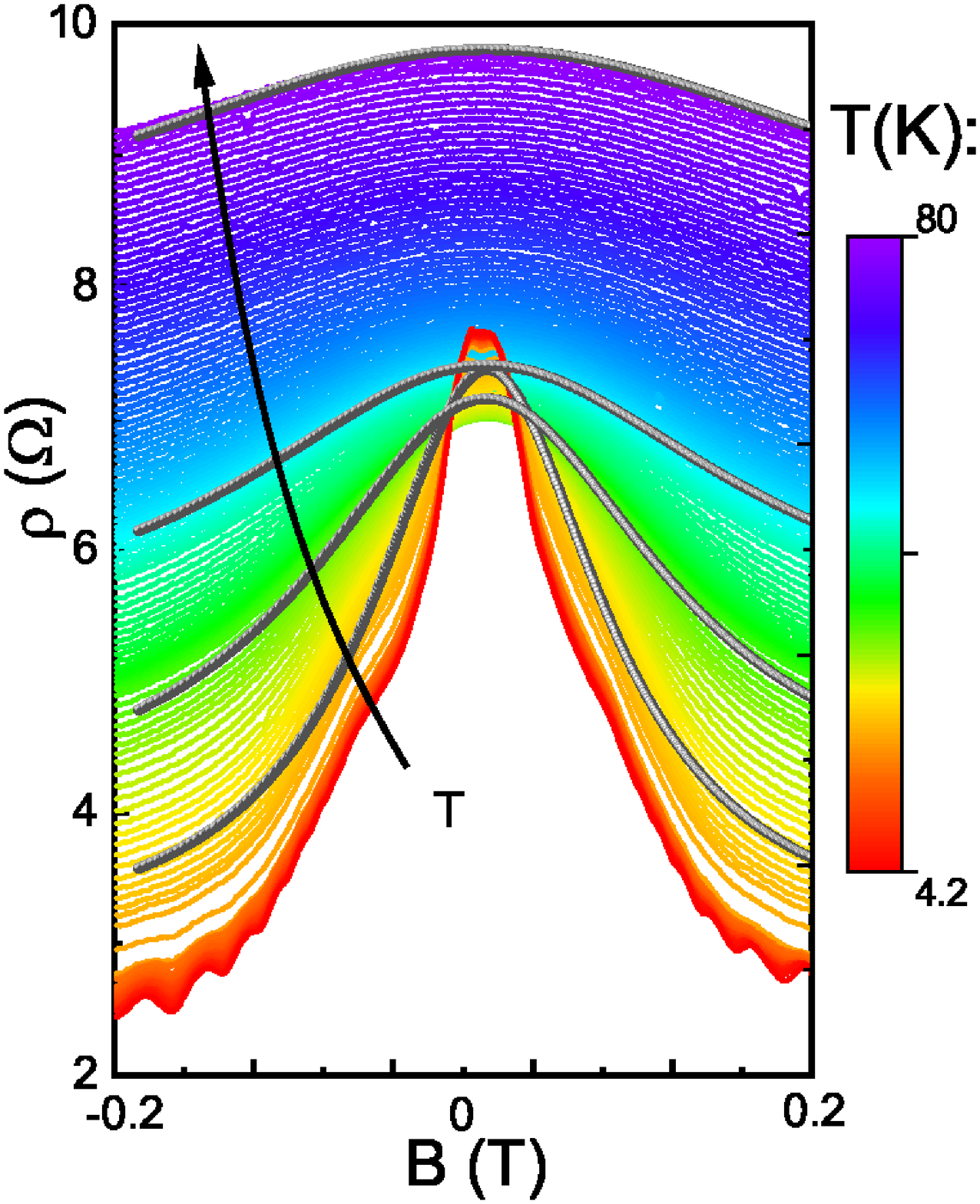}
\caption{(Color online)
Temperature-dependent  magnetoresistance  of  a mesoscopic GaAs bilayer for configuration C2.
 The circles are examples illustrating magnetoresistance calculated from
 Eq. (2) for different temperatures T(K): 15, 25, 43, 80.}
\end{figure}

 \begin{figure}[ht]
\includegraphics[width=9cm]{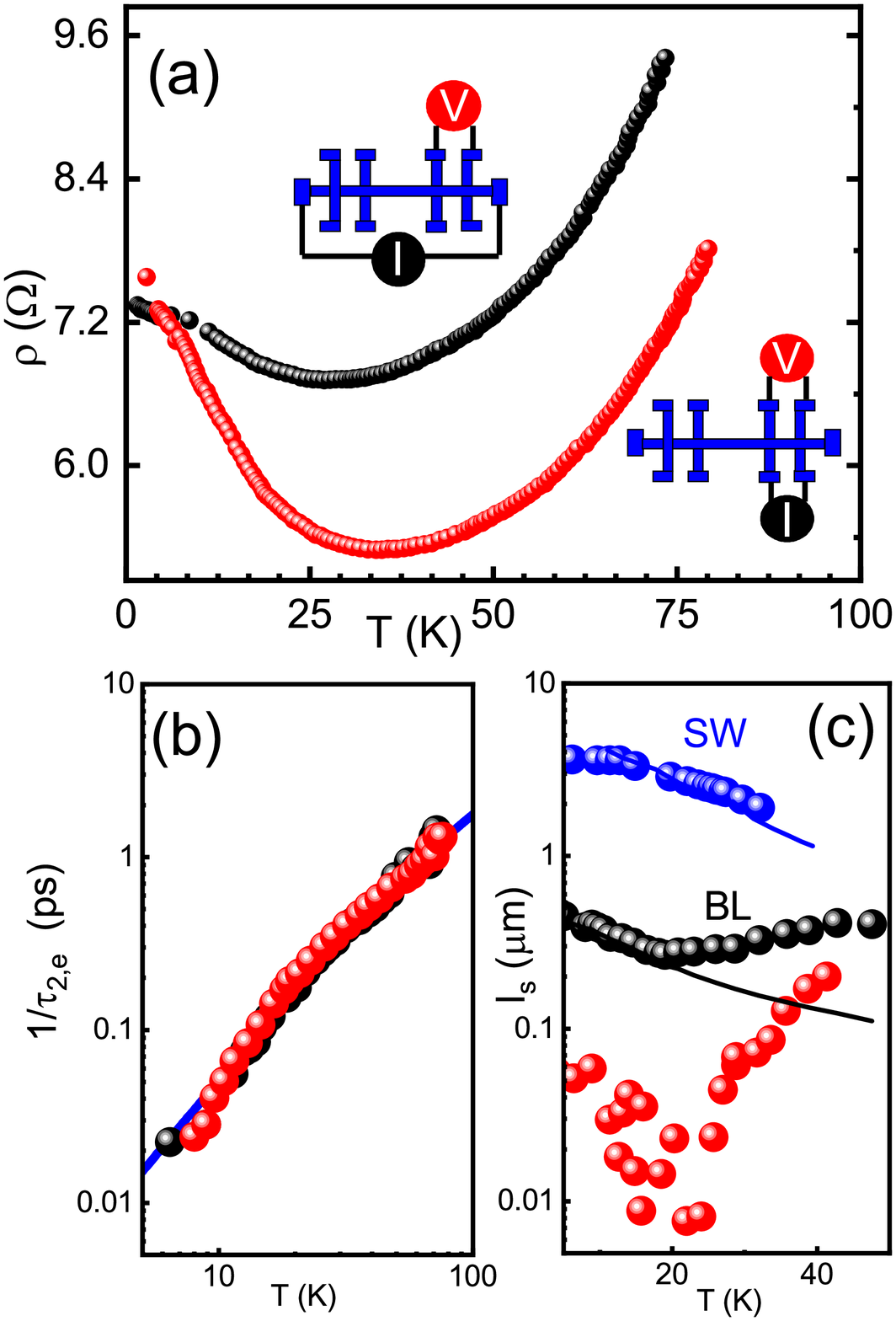}
\caption{(Color online)
(a) Temperature dependent resistivity of a GaAs bilayer in a Hall bar for different configurations in zero magnetic field.
(b) Relaxation rate $1/\tau_{2,e}$ as a function of temperature obtained by fitting the theory with experimental results.
Black circles - configuration C1, red circles - C2.
(c) Slip length as a function of the temperature for both
configurations. Thick blue line -the dependence $0.8l_{2,ee}$ (SW) and black line represents $0.5l_{2,ee}$ (BL). }
\end{figure}

 Fig. 5 shows a few representative curves for different current configurations. To underline the difference in the magnetoresistance shape
 for narrow and wide QWs, we plot the longitudinal magnetoresistance for a single well for the same configurations.
 When the current was applied between the source and the drain, the voltage was measured between the side probes (further referred to as conventional or C1 configuration),
 and the corresponding Lorentzian curve is wide and shows characteristic peaks in the region of small B in a single band well.
  When the current was applied between the side probes, the voltage was measured between the opposite side probes (further referred to as C2 configuration),
 the corresponding Lorentzian curve is narrow for both single and two subband wells. The characteristic peaks at small B tend to disappear for C2 configuration
 in both systems. The local minimum at B=0 is attributed to the classical size effect, and the observed weakening with rising temperature and configuration change was attributed to electron-electron scattering \cite{raichev}.

 A careful inspection of the magnetoresistance curves for bilayer samples reveals that the curves' shape is distorted with respect to a Lorentzian curve
 and shows a small shoulder near $B\approx 0.1 T$. This  weak  modification  of  the Lorentz shape could be attributed to the size effect \cite{raichev}.
 However, it is not clear why this effect is more pronounced in the bilayer in comparison with a single well. The shoulder disappears at $T>10 K$
 because the classical size effect is smeared out by temperature. In order to emphasize the hydrodynamic effects, we focus on high T measurements.

 Figures 6 and 7  show the evolution of longitudinal magnetoresistance with temperature for configurations C1 and C2, consecutively.
 The temperature increase leads to broader peaks and,  consequently, to a better agreement  with  Lorentzian shaped curves  at  high T.
 Remarkably, the zero-B peak decreases with T.  As  the  electron-electron  scattering  increases,  it is expected that the electron  system
 shifts  towards  the  hydrodynamic  regime, where  the  Gurzhi  effect  \cite{gurzhi}  is  possible.

It is worth noting that it  is  necessary  to not  only  understand the role of viscous effects in
 a clean electron system, but also to resolve the ballistic regime in the presence of a confining channel, impurities, and electron-electron scattering.
 The analytical solutions for transport equations in the ballistic regime have been obtained in model \cite{alekseev2, alekseev3}.
The  relaxation-time approximation for the electron-electron collision integral allows for either an analytical or a numerical solution
 of the kinetic equation \cite{dejong}, \cite{govorov}, \cite{scaffidi},
\cite{lucas1}, \cite{lucas2}, \cite{alekseev2}, \cite{alekseev3}, \cite{chandra},
\cite{holder}, however, in the presence of a magnetic field, the problem still remains complicated.

Fig. 8a demonstrates the temperature dependence of zero-B peak resistance  for a GaAs bilayer. One can see that the resistance
decreases with T for both current configurations, but the minimum in the $\rho(T)$ curve for configuration C2 is more pronounced.
We attribute this behaviour to a manifestation of the Gurzhi effect \cite{gurzhi}.

\section{Theory and discussion}
Application of the methods developed in the model \cite{raichev} to a two-subband system is a very challenging task,
and has not yet been established theoretically. Since the energy separation $\Delta$ in our bilayer is much smaller
than Fermi energy, we propose that the electron-electron scattering is essentially the same in both subbands.
In a two subband system $\rho_{total}=(\rho_{1}^{-1}+ \rho_{2}^{-1})^{-1}$,  and $\rho_{1}= m/e^{2}n_{1}\tau_{1}$ , $\rho_{2}= m/e^{2}n_{2}\tau_{2}$,
where  $1/\tau_{i}$ is the scattering  rate  which  includes both intrasubband and intersubband scattering and $m_{i}$, $n_{i}$ is the effective mass
and the density for the $i th$ subbands, respectively. For simplicity, we propose $m_{1}\approx m_{2}$,  $n_{1}\approx n_{2}$, $1/\tau_{1}\approx 1/\tau_{2}\approx 1/\tau $,
$\rho_{total}\approx m/e^{2}n_{total}\tau$.

Below we apply the models \cite{pellegrino, alekseev1}, because it captures all major hydrodynamic features, including the subtle effects related to
electron-electron scattering temperature dependence \cite{alekseev4}. The resulting resistivity of a 2D system in constrained geometry is given by

\begin{eqnarray}
\rho(T)=\rho_{0}\frac{1}{1-2\frac{D}{\xi W}\tanh(\frac{W}{2D})}
\end{eqnarray}

where $\sigma_{0}=e^{2}n\tau /m=1/\rho_{0}$ is the Drude conductivity, $\tau$ is momentum relaxation time due to interaction with phonons and
static defects, $D=\sqrt{\eta \tau_{2, ee}}$, $\xi=l_{s}\sinh(W/2D)+D\cosh(W/2D)$ is characteristic length
which depends on the boundary slip length $l_{s}$.
It has been shown that the equation for resistivity in zero magnetic field can be reduced to
 \cite{alekseev1, pellegrino, gromov}: $\rho(T) \approx \rho_{0}\left(1+\frac{\tau}{\tau^{*}}\right)$,
and that, in these conditions, the viscosity effect is regarded as a parallel channel of electron momentum relaxation with the characteristic time $\tau^{*}(\eta, l_{s})$.
This approach allows the introduction of the magnetic field dependent viscosity tensor and the derivation of the magnetoresisivity tensor \cite{alekseev1}:
\begin{eqnarray}
\rho_{xx}= \rho_{total}\left(1+\frac{\tau}{\tau^{*}}\frac{1}{1+(2\omega_{c}\tau_{2,ee})^{2}}\right),\,\,\,
\end{eqnarray}
where,  $\tau^{*}\approx\frac{W(W+6l_{s})}{12\eta}$, viscosity $\eta=\frac{1}{4}v_{F}^{2}\tau_{2,ee}$.
The relaxation rate $\frac{1}{\tau_{2}}$ relating to the process responsible for relaxation of the second moment
of the distribution function, such as  the head-on collisions of pairs of
quasi-particles and  scattering by static defect, gives rise to viscosity \cite{alekseev1, alekseev4}.

The momentum relaxation rate is expressed as $\frac{1}{\tau}=A_{ph}T+\frac{1}{\tau_{0, imp}}$,
where $A_{ph}$ is the term responsible for phonon scattering, and $\tau_{0, imp}$ is the scattering time due to static disorder
(not related to the second moment relaxation time) \cite{alekseev4}. Slip length is proportional to the electron - electron scattering length: $l_{s}=\alpha l_{e}$, where coefficient $\alpha$ depends
on the boundary roughness parameters \cite{kiselev}.

We fit the resistance in zero magnetic field with the  fitting parameters:
$\tau(T)$, $l_{s}(T)$. The magnetoresistance is fitted by Lorentzian curves (Eq. 2) with adjustable parameter  $\tau_{2,ee}$.

It is worth comparing the results obtained in a single subband system with those of a bilayer because, in a wide well, we
observe the Gurzhi effect in both configurations, while in a narrow well, R(T) decreases with T
only for the C2 arrangement \cite{gusev1}. Table 2 shows the parameters extracted
from the comparison of Eq. 2 and the experiment for bilayer (BL) and single well systems.

\begin{table}[ht]
\caption{\label{tab1} Fitting parameters of the electron system for a single subband well (SW) and a bilayer (BL) mesoscopic system for configuration C1, $T=14K$.
Parameters are defined in the text.}
\begin{ruledtabular}
\begin{tabular}{lcccccc}
&Well width &$\tau_{2,ee}$&$\tau_{2,imp}$ & $\tau$ & $\tau^{*}$ & Properties  \\
&($nm$) & $(10^{-12} s)$ & $(10^{-12} s)$ &  $(10^{-11} s)$ &  $(10^{-11} s)$ &   \\
\hline
&$16 $& $5.9$  & $6.9$ & $6.3$ &  $4$ & SW  \\
&$46$& $2.35$  & $4.4$ & $15$ &  $8.8$ & BL \\
\end{tabular}
\end{ruledtabular}
\end{table}

A notable distinction between a single and a two subband electron system is the presence  of additional
intersubband scattering in the BL. Electron-electron scattering in coupled quantum wells has been considered in ref. \cite{slutzky}.

The total inelastic scattering rate results from the intersubband transitions and that coming from the intrasubband processes,
$\left(\frac{1}{\tau_{ee}}\right)^{tot,i}= \left(\frac{1}{\tau_{ee}}\right)^{inter,i}+\left(\frac{1}{\tau_{ee}}\right)^{intra,i}$,
where $i=1,2$ is the subband number. One expects that the electron-electron scattering is more intensive
because the screening is more effective and because of the doubling of the  phase space for the intrasubband rate in comparison
with the single-band one \cite{slutzky}.

 The inelastic scattering rate for  the intrasubband processes
is given by:

\begin{eqnarray}
\left(\frac{\hbar}{\tau_{ee}}\right)^{intra,i}= -A_{1}\frac{(kT)^2}{E_{F}}+A_{2}\frac{(kT)^2}{E_{F}}ln\left(\frac{4E_{F}}{kT}\right)
\end{eqnarray}

And for intersubband scattering:

\begin{eqnarray}
\left(\frac{\hbar}{\tau_{ee}}\right)^{inter,i}= -B_{1}\frac{(kT)^2}{E_{F}}
+B_{2}\frac{(kT)^2}{E_{F}}ln\left(\frac{4E_{F}}{\Delta}\right) \nonumber \\
+B_{3}\frac{(kT)^2}{E_{F}}ln\left(\frac{\Delta}{kT}\right)
\end{eqnarray}
Everywhere $A_{i}$ and $B_{i}$ are positive numerical constants of order unity with $B_{2}> B_{3}$. Indeed
the two rates are almost equal $\left(\frac{1}{\tau_{ee}}\right)^{tot,1}= \left(\frac{1}{\tau_{ee}}\right)^{tot,2}$.

Inter electron collisions, given by Equations (2) and (3), are an oversimplification of perturbation in a Fermi gas
considered in all theoretical models \cite{gurzhi}-\cite{lucas1}.  Stress relaxation rate may have slightly different
parameters and logarithmic temperature behaviour \cite{alekseev1}.  However, it is useful to compare our results with the existing theory.
We compare the temperature dependence of  $\frac{1}{\tau_{2,e}(T)}=\frac{1}{\tau_{2,ee}(T)}-\frac{1}{\tau_{2,imp}}$ with Equations (3) and (4), as shown in Fig 8b.
The following parameters
are extracted: $\tau_{2,imp}=3.1\times10^{-12}$ s, for C1, and $\tau_{2,imp}=4.2\times10^{-12}$ s, for C2. In the single well, we compare the results for relaxation rate
with $A_{ee}\frac{(kT)^{2}}{\hbar E_{F}}$, and find parameter $A_{ee}=0.2$ \cite{raichev}. Figure 8b shows the theoretical predictions for parameters $A_{i}=B_{i}=0.12$.

More intensive e-e scattering may lead to diffusive boundary scattering and improve the hydrodynamic regime. The Poiseuille type flow
 supports  a very small slip length if compared to the characteristic size of the system. Figure 8c shows the slip length as a function of temperature extracted from the fit with experimental results. One can see that $l_{s}$ for configuration C2 roughly
follows T-dependence for $l_{2,ee}$ below 25 K. The black line represents curve $0.5l_{2,ee}$, which corresponds to conditions for diffusive boundary scattering.
At higher temperatures the deviation requires additional explanations. We find that $l_{s}$ for configuration C1 is
negligibly small and  can not be reliably extracted from comparison with the theory. Figure 8c shows also the slip length for a single well.
One can see that the extracted $l_{s}$ in a SW  is longer than $l_{s}$ in a BL and follows to dependence $0.8l_{2,ee}$ .
For shorter slip length, the velocity profile becomes more parabolic, as is shown in Figure 1.
Let us compare the second (hydrodynamic) term in Equation (2) for both SW and BL systems at B=0. The ratio is given by

\begin{eqnarray}
\frac{\rho_{h}^{BL}}{\rho_{h}^{SW}}\approx 6\left(\frac{\tau^{BL}}{\tau^{SW}}\right)\left(\frac{\tau_{2,e}^{BL}}{\tau_{2,e}^{SW}}\right)\left(\frac{v_{F}^{BL}}{v_{F}^{SW}}\right)^{2}\left(\frac{l_{s}^{SW}}{W}\right)
\end{eqnarray}

 where we consider that $6l_{s}^{BL}/W\ll 1$ and $6l_{s}^{SW}/W\gg 1$. In a real system, the parameter $\tau(T)$ in a mesoscopic sample ($\tau_{mesosc}$) was found to be larger than in a macroscopic structure ($\tau_{macrosc}$). The same tendency has been observed in SL. For example we obtain the ratio $\tau_{mesosc}/\tau_{macrosc}=2.9$ for a BL and $\tau_{mesosc}/\tau_{macrosc}=1.3$ for a SW at 4.2 K. The distance between probes and the width of the sample was much smaller than the ballistic mean free path, and a direct comparison and interpretation of the data extracted from the macroscopic and mesoscopic samples is not well established. In the hydrodynamic approximation, $\tau$ describes relaxation of all angular harmonics of the distribution function except the zero one (ref.\cite{alekseev1}). The introduction of the unified times for all harmonics is a crude approximation, but it simplifies the solution of the kinetic equation.
For parameters $\tau^{BL}/\tau^{SW}\approx 2$, $\tau_{2,e}^{BL}/\tau_{2,e}^{SW}=0.5$
 and  $v_{F}^{BL}/v_{F}^{SW}\approx 0.7$ we obtain $\frac{\rho_{h}^{BL}}{\rho_{h}^{SW}}\approx  1.8$. Therefore, the less viscous BL system becomes more favorable
  to following a Poiseuille type flow when $l_{s}^{SW}\sim W$.

We conclude that the two subband system in general offers stronger
viscous flow effects. First, this is because the hydrodynamic effect can be realized in a BL in a wider temperature
range in comparison with a SW. As was indicated in the introduction, the hydrodynamic regime requires
$l/l_{e} \gg  1$ and  $l_{e}/W\ll 1$. In a BL with the same total density, we obtain $l_{e}/W\approx
0.1 $ at T=4.2 K, while in a single subband system the ratio $l_{e}/W\approx 0.2 $  is reached at T=30 K. Second,
the other key physical parameter controlling the Poiseuille-like flow profile, the slip length $l_{s}$, is smaller in a
BL. The slip length can be controlled by e-e scattering, which is more intensive in a BL due to the doubling of the phase space and intersubband
scattering. The hydrodynamic regime requires $W/l_{s} \gg 1$, which  indeed is realized in a BL at low temperature T=4.2K, while in a SW
$W/l_{s} =2$, even at high temperatures.

It is worth noting that the advantage of a BL in comparison with a SW with the same total density is smaller kinematic viscosity $\nu$.
It allows the achievement of  large Reynolds numbers for relatively small injected current in order to observe electronic pre-turbulence
in a mesoscopic scale \cite{gabbana}.

\section{Summary and conclusion}
In conclusion, we report the appearance of the Gurzhi effect in a mesoscopic two-dimensional electron system in a wide GaAs
quantum well with two occupied subbands. We observe a large negative magnetoresistance
in a wide temperature range. By comparing theory and the experiment, we determine the characteristic relaxation  time
of electrons caused by electron-electron scattering. In addition we demonstrate that the slip length is shorter in a BL,
which results in a more parabolic flow profile in comparison with a SW.

\section{Acknowledgments}
 The financial support of this work by RSF Grant No. 21-12-00159, São Paulo Research Foundation (FAPESP) Grants No. 2015/16191-5 and No. 2017/21340-5, and the National Council for Scientific and Technological Development (CNPq) is acknowledged. We  thank  O.E. Raichev  for  the  helpful  discussions.

\end{document}